# Cascaded Structural Luneburg Lens for Broadband Cloaking and Wave Guiding

Liuxian Zhao[a], Miao Yu[a,b,*]

[a] Institute for Systems Research, University of Maryland, College Park, MD, 20742, USA

[b] Department of Mechanical Engineering, University of Maryland, College Park, Maryland 20742, USA

*Author to whom correspondence should be addressed: mmyu@umd.edu

ABSTRACT

In this paper, we explore the concept of structural Luneburg Lens (SLL) as a design framework for performing dynamic structural tailoring to obtain a structural wave cloak and a structural waveguide. The SLL is a graded refractive index lens, which is realized by using a variable thickness structure defined in a thin plate. Due to the thickness variation of the plate, the refractive index decreases radially from the centre to the outer surface of the lens. By taking advantage of the unique capabilities of SLL for flexural wave focusing and collimation, we develop a structural wave cloak and waveguide based on cascaded SLLs. The cascaded SLL design enables the integration of functional devices into thin-walled structures while preserving the structural characteristics. Analytical, numerical, and experimental studies are carried out to characterize the performance of the SLL cloak and the SLL waveguide. The results demonstrate that that these cascaded SLL devices exhibit excellent performance for structural wave cloaking and waveguiding over a broadband operating frequency range.

# 1. Introduction

Manipulation of flexural wave propagation in thin plates has received a growing interest. Many efforts on flexural wave manipulation have been reported, including flexural wave focusing, wave guiding, and cloaking [1-8]. Current approaches for focusing and guiding of flexural waves include: i) varying the effective refractive index profile, ii) exploiting frequency bandgaps, and iii) employing topological techniques [9, 10]. In many of these approaches, wave manipulation is achieved by using periodic structures, such as phononic crystals and metamaterials, whose refractive indices and frequency bandgaps can be tailored via the design of unit cells. For example, Sun *et. al.* [11] developed phononic crystal plate waveguides consisting of circular steel cylinders, in which wave propagation is well confined. Darabi *et. al.* [12] combined metamaterials with piezoelectric transducers to achieve focusing and guiding of flexural waves. However, these structures are based on Bragg scattering bandgaps, which are narrowband in nature. Structures with a continuous refractive index profile have also been explored, which can help mitigate the aberration and frequency dependence issues associated with discrete structures. For example, Zareei *et. al.* [7] designed a hyperbolic secant lens with a variable thickness profile in a thin plate to focus and guide the flexural wave propagation. However, the residual thickness at the lens's centre is only ¼ of the plate thickness, which could compromise the rigidity of the plate.

On the other hand, there is a great need to protect objects from unwanted vibrations and incident waves, including undesirable sound and elastic waves (e.g., acoustic noise, shock waves, and seismic waves) [13, 14], unwanted mechanical resonances (vibration isolation) [15-17], and undesired reflections from bolted joints or holes in structural health monitoring [18, 19]. Current techniques to address this need include strengthening structures, modifying structural resonances through additive engineering or anti-resonance, developing flexible structures that can withstand large deformations, and adding dissipative elements [20-24].

Recently, phononics crystals and acoustic metamaterials have been exploited for vibration isolation, earthquake engineering, and flexural wave cloaking. Most of these efforts have been focused on the development of periodic cloaking structures or transformation acoustic approaches [25-34]. However, periodic structures have the limitation of narrow band [35, 36] and the transformation acoustics method requires complicated design and stringent refractive indices [37, 38].

In this work, we propose to use a structural Luneburg lens (SLL) based on a variable thickness structure defined in a thin plate as a framework for structural wave guiding and cloaking. Similar to its counterpart of optical and acoustic Luneburg lenses, SLL has a gradient refractive index profile that decreases radially from the lens centre to its outer surface. Optical Luneburg lens [39-41] has been well studied and its focusing and collimation capabilities have been applied to many applications, such as communications and nuclear scattering [42-44]. In recent years, acoustic Luneburg lens has been developed for omni-directional acoustic wave focusing and collimation by using phononic crystals and acoustic metamaterials [45-50]. More recently, structural Luneburg lens has also been investigated. Torrent *et. al.* [51] studied different omnidirectional refractive devices including Luneburg lens based on graded phononic crystals for flexural wave manipulation. Tol *et. al.* [52] explored a phononic crystal Luneburg lens based on a hexagonal unit cell with blind holes for elastic wave focusing. In addition, Climente *et. al.* [53] explored several gradient index lenses including Luneburg lens based on thickness variations of a thin plate for broadband flexural wave manipulation.

Here, we apply the concept of SLL to achieve cascaded SLL devices for structural cloaking and wave guiding. Our lens has the following attributes: i) Since the lens design does not require bandgaps and is based on a continuous graded structure, it is broadband, ii) the lens in theory is capable of zero-aberration structural wave guiding and cloaking, iii) the minimum thickness of our lens is ½ of the constant plate thickness, which reduces the risk of plate damage.

## 2. Cascaded Structural Luneburg Lens Design

Figure 1 (a) shows the cross-section view of a single SLL based on a variable thickness structure defined in a thin plate. The lens is defined by a circular region with $r \leq R$, where $r$ is the radial distance from the central point and $R$ is the radius of the variable thickness region. For the SLL shown in Figure 1, based on the optical Luneburg lens principle [54, 55], the radially gradient refractive index satisfies the following:

$$n(r) = \frac{\sqrt{2R^2 - r^2}}{R}, \tag{1}$$

As a flexural wave propagates through the SLL, the phase velocity is given as [56]:

$$c = \frac{\omega}{k} = \left(\frac{Eh^2\omega^2}{12\rho(1-v^2)}\right)^{\frac{1}{4}}, \tag{2}$$

where $\omega$ is the angular frequency, $k$ is the wavenumber, $E$ is the Young's modulus, $\rho$ is the density, and $v$ is the Poisson's ratio.

Based on Equations (1) and (2), the variable thickness in the lens region is derived as:

$$h(r) = \frac{h_0 R^2}{2R^2 - r^2}, \tag{3}$$

where $h_0$ is the constant thickness of plate.

For proof-of-concept, we consider a SSL with a radius $R = 0.04$ m in a plate with a constant thickness $h_0 = 0.004$ m, which is used as a basis to construct the cascaded SLLs as the structural wave cloak and waveguide. The refractive index and variable thickness along the radial distance are plotted in Figure 1 (b). It can be seen that the maximum refractive index reaches 1.42 at the lens center, which requires the minimum residual thickness to be $h_0/2$.

Two cascaded SLLs are used to construct a structural cloaking device, as shown in Figure 1 (c). When a line source is used for excitation, the flexural wave passing through the first SLL generates a focus, and then the wave is collimated again after passing through the second SLL. The incident flexural wave bypasses the region between the junctions of the two SLLs, rendering a "cloaking zone". Hence, for an object placed in the cloaking zone, direct interaction of the object with the propagating flexural wave is isolated. Furthermore, four cascaded SLLs are used to construct a structural waveguide, as shown in Figure 1 (d). Consider a point source excitation on the edge of the first SLL, the flexural wave is collimated after propagating through the first SLL, and then focused again after passing through the second SLL. The same process is repeated after the wave passes the third and fourth SLLs, producing a circular wavefront at the exit of the four cascaded SLLs without much loss. Therefore, the four (or any other even numbers above 4) cascaded SLLs can serve as a waveguide for structural waves.

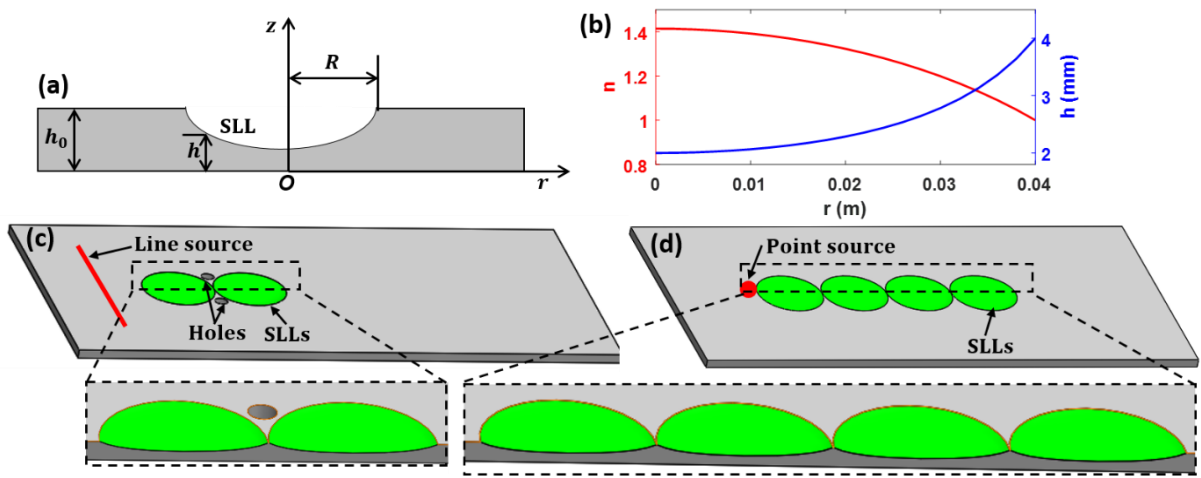

**Figure 1: Cloaking and wave guiding based on cascaded structural Luneburg lens and their design principles. (a) Schematic of SLL based on variable thickness structure defined in a thin plate for manipulating flexural waves. (b) Refractive index ($n$) and variable thickness ($h$) as a function of radial location $r$. (c) Two cascaded SLLs for structural cloaking. (d) Four cascaded SLLs as structural waveguide. The insets in (c)**

**and (d) are the close-up cross sections of the cascaded SLLs. The green coloured regions in (c) and (d) represent the SLLs.**

## 3. Analytical and Numerical Studies

Flexural wave propagation through two-dimensional SLLs was analysed by using a geometrical acoustics approach in Hamiltonian formulation [57, 58]. This approach offers a simple solution to analyse the flexural wave propagation in plates of variable thicknesses, including plates with a cylindrically symmetrical profile, in which there is no geometrically induced anisotropy of flexural wave velocity. In this work, the ray trajectory method was used to study flexural wave focusing and collimation by using SLLs. Consider a bundle of incident flexural rays (with a width of 3$R$/2) propagate along the $x$ direction of the constant thickness region of the plate and interact with the cascaded SLLs. The ray trajectories of flexural wave propagation are illustrated in Figure 2(a). By combining the focusing and collimating properties of two SLLs, "ray-free" regions can be created, which can be used as "cloaking zone.

We further conducted numerical simulations to investigate a SLL structural cloaking device defined in a thin aluminium plate (material properties: density $\rho = 2700$ kg/m$^3$, Young's modulus $E = 70$ GPa, and Poisson's ratio $v = 0.33$). The dimensions of the plate are $l_0 \times w_0 \times h_0 = 0.45$ m $\times$ 0.15 m$\times$ 0.004 m. The radius of the Luneburg lens is $R = 0.04$ m. Two circular holes with a radius of $a = 0.008$ m were included in the cloaking zone. Full 3D wave propagation simulations in both frequency and time domains were performed by using commercial software COMSOL. To reduce the boundary reflections, Perfectly Matched Layers (PML) were applied to the four sides of the plate in the frequency domain, and Low Reflecting Boundaries (LRB) were used in the time domain analysis.

A line source with a frequency range of 25 – 70 kHz was used for excitation. Note that the lower bound frequency $f = 25$ kHz was determined according to the scattering regime (, where is the wavelength). The upper bound frequency $f$ = 70 kHz was chosen due to the limitation of numerical simulations, which require 5-10 elements per wavelength. The displacement field and phase distributions of flexural wave propagation for $f$ = 40 kHz are shown in Figure 2(b) and Figure 2 (c), respectively. It can be clearly seen that the wave focused at the interface between the two SLLs, and propagated as a plane wave after passing through the second SLL. Furthermore, in the cloaking zone where the two holes were located, the scattering was substantially suppressed and the waveform and phase distributions remained little distorted after passing through the two lenses (Figure 2(b) and (c)). In order to investigate the broadband characteristic of the cloaking device, phase profiles at a representative position (white dashed line in Figure 2(c)) were calculated for different frequencies, as shown in Figure 2(d). The phase profiles were almost constant over a broadband frequency range of 25 – 70 kHz, demonstrating the broadband cloaking ability. The discontinuous phase profiles observed at high frequencies are believed to be due to the limitation of mesh size used in the simulations.

By using the above mentioned analytical and numerical methods, the SLL based waveguide was also explored. The ray trajectories of flexural wave propagating and interacting with the SLL based waveguide were obtained, as shown in Figure 2 (e). A point source on the edge of the first SLL excited a family of incident rays, which travelled through four cascaded SLLs and experience alternating collimation and focusing before exiting the waveguide with circular waveforms. In principle, the SLL waveguide allows flexural wave propagation over a long distance without much attenuation.

In the numerical simulations conducted for the SLL based waveguide, a point source with a frequency range of 25 – 70 kHz was used for excitation. Figure 3 (f) shows the waveform distribution for $f$ = 40 kHz when the flexural wave interacts with the SLL waveguide. It can be observed that the flexural wave was alternately collimated and focused by the four cascaded SLLs, and eventually exited the waveguide with a circular waveform that resembled the original point source. Similar to cloaking, the SLL waveguide also exhibited a good broadband performance over the frequency range of 25 – 70 kHz, as shown in Figure 2(g). For comparison, flexural wave propagation in a similar plate with a constant thickness (without SLL cloak or waveguide) was also studied in numerical simulations, as shown in Figure S1 of the supplementary information. For the SLL cloak, the vibration displacement in the cloaking zone is shown to be greatly reduced over a broadband frequency range of 25 – 70 kHz as compared with the case without the SLLs. At 47 kHz, a minimum amplitude with 61% reduction was achieved (Figure S1(c)). For the SLL waveguide, the vibration displacement was enhanced in a broadband frequency range from 25 – 70 kHz with a gain ranging from 3 to 5 (maximum gain of 5 at 47 kHz) (Figure S1(f)). Furthermore, a transmission efficiency of 0.76 to 0.93 over the frequency range of 25 – 70 kHz can be obtained (Figure S1(g)).

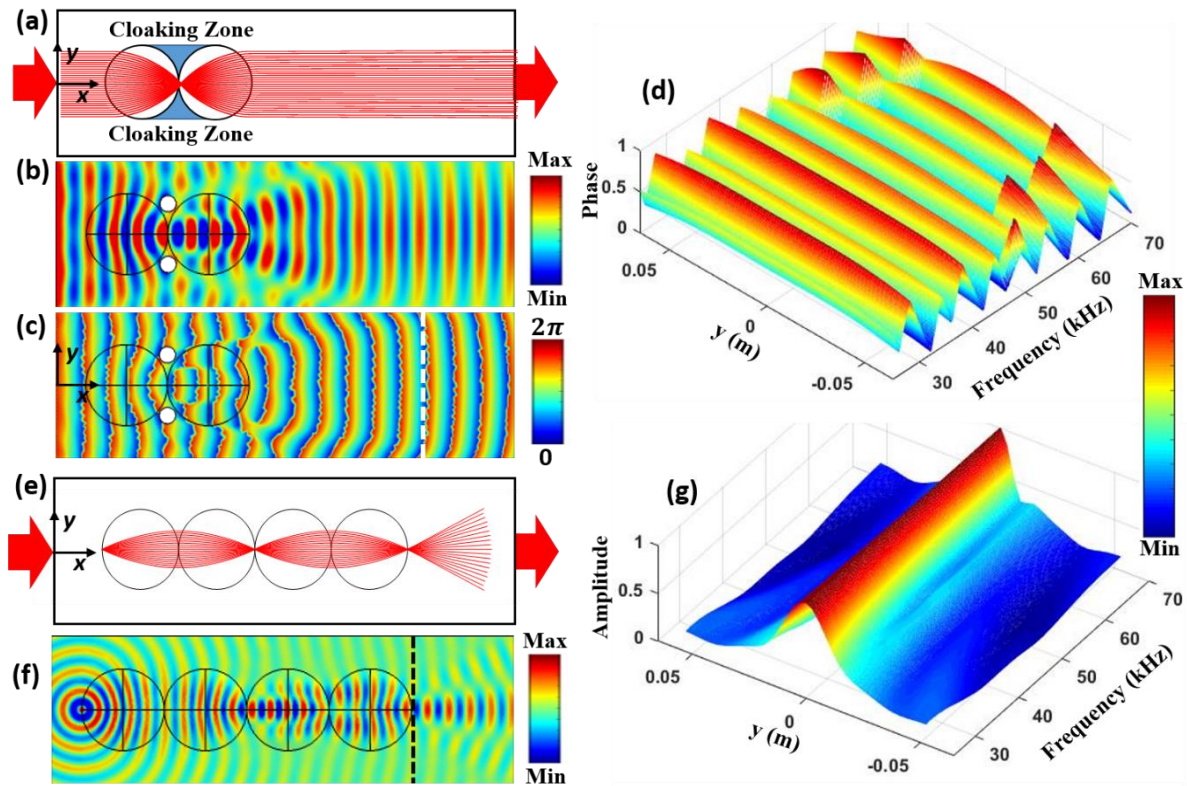

**Figure 2: Analytical and numerical results for SLL based structural wave cloak and waveguide. Ray trajectories of (a) structural wave cloak with two cascaded SLLs and (e)**

**structural waveguide with four cascaded SLLs. Numerical simulations of (b) waveform and (c) phase distribution for SLL based cloak at $f$ = 40 kHz. (d) Phase distribution over the frequency range of 25 - 70 kHz along the white dashed line in (c). (f) Numerical simulations of waveform for SLL based waveguide at $f$ = 40 kHz. (g) Normalized amplitude distribution along the black dashed line in (f) over the frequency range of 25 - 70 kHz. The normalization is based on the maximum value along the $y$ axis for each frequency.**

Furthermore, time domain analysis was performed numerically to investigate the cloaking and wave guiding with cascaded SLLs. A line source and a point source with a signal of 3-count Hanning-windowed tone burst at 40 kHz were used for excitation for the SLL cloak and the waveguide, respectively, as shown in Figure S2.

For SLL cloaking, the obtained waveforms at different time steps of $t$ = 0.04, 0.07, 0.09, and 0.25 ms are shown in Figure 3 (a)-(d), respectively. At $t$ = 0.04 ms, the generated flexural waves propagated forward with a plane wavefront. At $t$ = 0.07 ms, the flexural waves interacted with the Luneburg lens, the wavefront gradually became narrower. At $t$ = 0.09 ms, the flexural waves were localized and formed a focus at the interface between the two SLLs. At $t$ = 0.25 ms, the focused waves after interacting with the second SLL propagated forward as a plane wave.

The waveform obtained for flexural waves interacting with SLL waveguide at different times are shown in Figure 3 (e)-(j). At $t$ = 0.03 ms, the flexural waves generated from the point source propagated in all directions. At $t$ = 0.08 ms, after interacting with the first SLL, the waves became collimated. At $t$ = 0.13 ms, after interacting with the second SLL, the wavefront gradually became narrower and a focus was formed. After passing through the third SLL, the flexural waves became collimated again at $t$ = 0.18 ms, and then were focused on the edge of the fourth SLL at $t = 0.23$ ms. At $t = 0.25$ ms, the waves continued to propagate forward with a circular wavefront, demonstrating the cascaded SLLs as a waveguide.

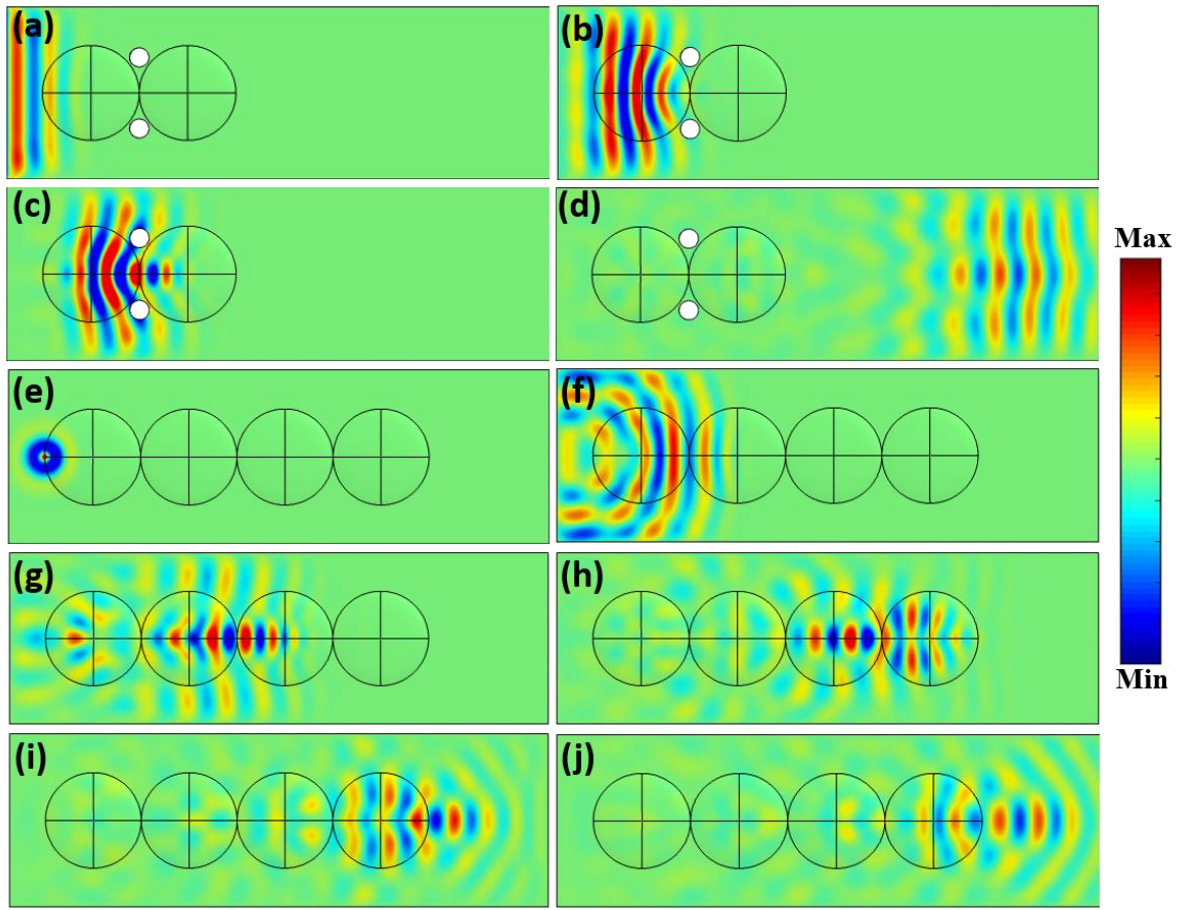

**Figure 3: Numerical simulations of transient responses obtained for structural wave cloak and structural waveguide. (a) - (d) are the flexural wave propagation through the SLL cloak at time instants of $t = 0.04$ ms, $t = 0.07$ ms, $t = 0.09$ ms, and $t = 0.25$ ms, respectively. (e) - (j) are the flexural wave propagation through the SLL waveguide at time instants of $t = 0.03$ ms, $t = 0.08$ ms, $t = 0.13$ ms, $t = 0.18$ ms, $t = 0.23$ ms, and $t = 0.25$ ms, respectively.**

## 4. Experimental Studies

Experimental studies were carried out to validate the SLL based structural wave cloak and waveguide. The experimental setup and the fabricated SLL cloak and waveguide are shown in Figure 4. The SLL cloak and waveguide were fabricated on a 6061-aluminium plate (from McMaster-Carr) with dimensions of $l \times w \times h = 0.6$ m $\times 0.3$ m $\times 0.004$ m. For structural cloaking, two holes with a radius of $a = 0.008$ m were drilled in the cloaking zone. The four sides of the plate were covered with damping clay to reduce the boundary reflections. The radius of each SLL was $R = 0.04$ m. Five rectangular piezoelectric transducers (dimensions of 20 mm $\times$ 15mm $\times$ 1 mm from STEMiNC Corp.) were attached to the plate surface to serve as a line source for the SLL cloak. A circular piezoelectric transducer (12 mm in diameter and 0.6 mm in thickness from STEMiNC Corp.) was used to generate a point source for the SLL waveguide. The piezoelectric transducers were bonded to the plate by using 2P-10 adhesive (from Fastcap, LLC). The two vertical sides of the plates were fixed on a frame, as shown in Figure 4 (a). Photographs of the fabricated SLL cloak and waveguide are shown in Figure 4 (b) and (c), respectively. A Polytec PSV-400 scanning laser vibrometer (SLV) were used to measure the propagating wave field by recording the out-of-plane component of the particle velocity on the plate over a scanned area (blue coloured area in Figure 4). The scanning was performed on the flat side of the plate. In the experiments, a signal of 3-count Hanning-windowed tone burst was used for excitation at the frequency of 40 kHz. Time domain responses were measured to validate the performance of both the SLL cloak and the SLL waveguide.

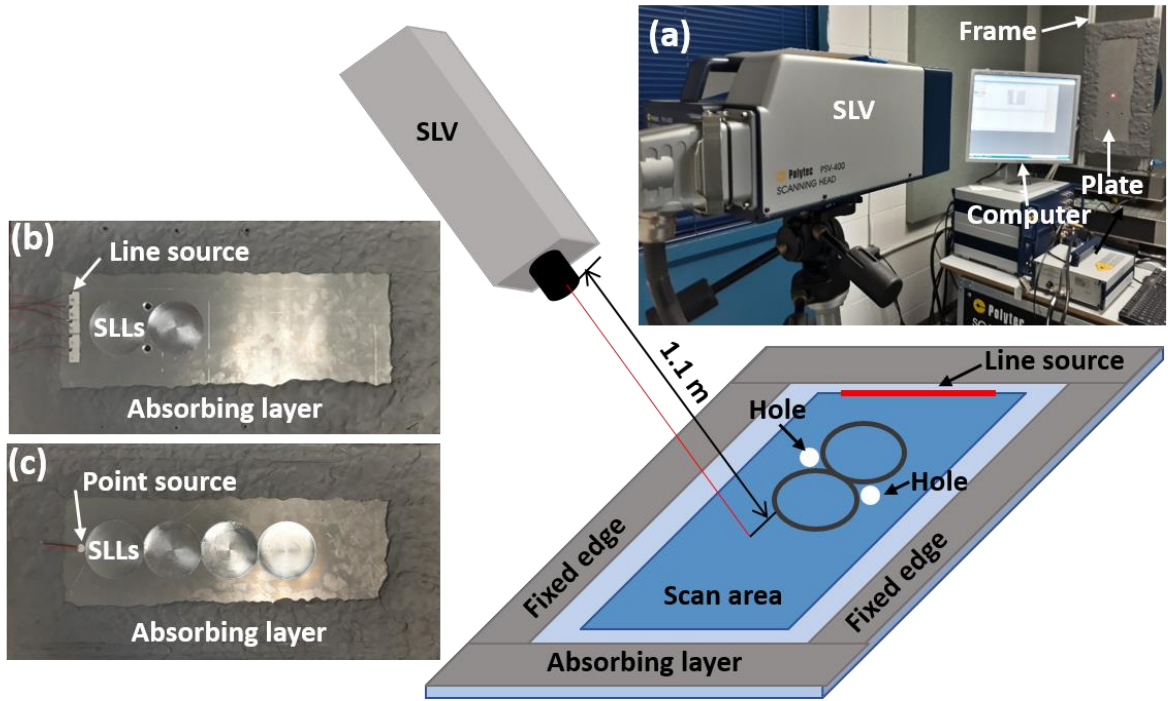

**Figure 4: Experimental setup and fabricated SLL cloak and waveguide. A scanning laser vibrometer was used to measure the full-field wave propagation in the scan area. (a) Photo of the experimental setup. The thin plate was constrained by two vertical fixtures and covered with absorbing layers. (b) Fabricated structural cloak with 2 cascaded SLLs defined in the thin plate. (c) Fabricated structural waveguide with 4 cascaded SLLs.**

The measured out-of-plane particle velocity waveforms of the wave propagation through the SLL cloak and the SLL waveguide are shown in Figure 5, which are in excellent agreement with the numerical simulation results (see Figure 3). As shown in Figure 5 (a)-(d), after propagating through the SLL cloak, the original plane wave was undisturbed even with the presence of two drilled holes. For the wave propagation through the SLL waveguide (Figure 5 (e)-(j)), the outgoing flexural wave at the exit of the waveguide exhibited the same circular wavefront as that of the original point source. These results experimentally validated the cloaking and wave guiding capacities of cascaded SLLs.

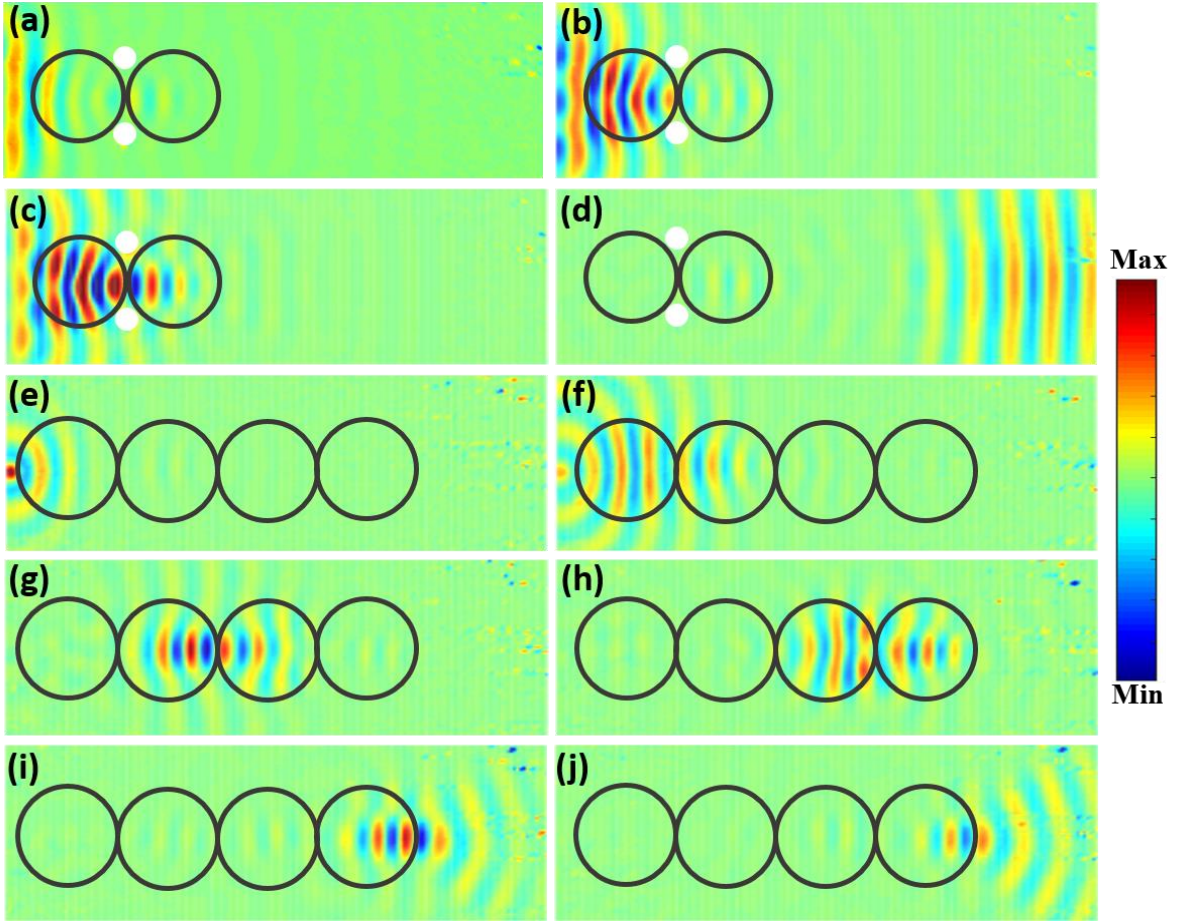

**Figure 5: Measured transient response for the structural wave cloak and waveguide.** (a) - (d) Flexural wave propagation through the structural wave cloak at time instants of $t = 0.04$ ms, $t = 0.07$ ms, $t = 0.09$ ms and $t = 0.25$ ms. (e) - (j) Flexural wave propagation through the structural waveguide at time instants of $t = 0.03$ ms, $t = 0.08$ ms, $t = 0.13$ ms, $t = 0.18$ ms, $t = 0.23$ ms, and $t = 0.25$ ms.

## 5. Conclusions

We analytically, numerically, and experimentally demonstrated for the first time that broadband structural wave cloaking and wave guiding can be achieved by using a set of structural Luneburg lenses. Each individual SLL is based on a variable thickness structure defined in a thin plate, which renders continuous gradient of refractive index along the radial direction. This SLL allows aberration-free structural wave focusing and collimation over a broad frequency range. By taking advantage of the properties of single SLL, a structural wave cloak and a structural waveguide were realized based on two and four cascaded SLLs, respectively. The ray trajectory method was used to calculate the wave propagation through the structural wave cloak and the waveguide, demonstrating their working principles. Furthermore, numerical simulations were carried out to characterize the performance of the structural wave cloak and the waveguide, which were validated by the experimental studies. Our results showed that the SLL cloak allowed flexural wave propagation without much disturbance by the objects present in the cloaking zone, while the SLL waveguide allowed flexural wave propagation over a distance without much attenuation and wavefront change. This work provides insights into the development of various functional devices based SLLs with unique flexural wave manipulation properties.

**Acknowledgements**

**Conflict of Interest**

The authors declare no conflict of interest.